\documentclass[prb]{revtex4}
\usepackage{graphicx}
\usepackage{amsmath}
\usepackage{amssymb}
\usepackage{latexsym}

\begin{document}
\title{
Scattering matrix approach to the resonant states and $Q$-values
of microdisk lasing cavities.}

\author{A. I. Rahachou and I. V. Zozoulenko}
\affiliation{Department of Science and Technology (ITN), Link\"oping University, 601 74
Norrk\"oping, Sweden}
\date{\today}

\begin{abstract}
We have developed a scattering-matrix approach for  numerical
calculation of resonant states and $Q$-values of a nonideal
optical disk cavity of an arbitrary shape and of an arbitrary
varying refraction index. The developed method has been applied to
study the effect of surface roughness and inhomogeneity of the
refraction index on $Q$-values of microdisk cavities for lasing
applications. We demonstrate that even small surface roughness
($\Delta r \lesssim\lambda/50$) can lead to a drastic degradation
of high-$Q$ cavity modes by many orders of magnitude. The results
of numerical simulation  are analyzed and explained in terms of
wave reflection at a curved dielectric interface combined with the
examination of Poincar\'{e} surfaces of section and Husimi
distributions.
\end{abstract}
\maketitle

\section{Introduction}

Dielectric and polymeric microcavities represent a great potential
for possible applications in lasing optoelecronic devices
\cite{Yamamoto,Nockel}. In conventional lasers, a significant
fraction of optical pump power is lost and a rather high threshold
power is needed to initiate the lasing effect. In contrast,
spherical and disk cavities can be used to support highly
efficient low-threshold lasing operation. The high efficiency of
such devices is related to the existence the natural cavity
resonances. These resonances are known as morphology-dependent
resonances or whispering gallery modes \cite{Hill_Benner}. The
nature of these resonances can be envisioned in a ray optic
picture, when light is trapped inside the cavity through the total
internal reflection on the cavity-air boundary.

In dielectric cavities optically pumped quantum wells, wires or
dots provide an active medium sustaining the lasing operation
\cite{Slusher_1992,Slusher_1993,Fujita,Gayral,Seassal}. Polymeric
microcavity lasers are made with an active medium including host
and guest molecules \cite{Berggren,Inganas,Polson2002}. The
absorbed light is transferred from the photoexcited host molecules
in the non-radiative way by means of resonant energy transfer to
the guest molecules. A stimulated emission from the active medium
of dielectric and polymeric cavities is trapped in high-$Q$ modes
for a very long time. This leads to a significant increase of
intensity of radiation inside the cavity and hence to
low-threshold laser operation.

One of the most important characteristics of cavity resonances is
their quality factor ($Q$-factor) defined as $Q=2\pi*$(Stored
energy)/(Energy lost per cycle). The high value of the $Q-$factor
results from very low radiative losses that are mainly caused by
radiation leakage due to diffraction on the curved interface. An
estimation of the $Q$-factor in an ideal circular disk cavity of a
typical diameter $d\sim 10\mu$m for a typical WG resonance gives
$Q\sim 10^{13}$ (see below, Eq. (\ref{Eq_T})). At the same time,
experimental measured values reported so far are typically in the
range of $10^3\sim 10^4$ or lower
\cite{Slusher_1992,Slusher_1993,Fujita,Gayral,Seassal,Berggren,Inganas,Polson2002}.
A reduction of a $Q$-factor may be attributed to a variety of
reasons including side wall geometrical imperfections,
inhomogeneity of the diffraction index of the disk, effects of
coupling to the substrate or pedestal and others. A detailed study
of the effects of the above factors on the characteristics and
performance of the microcavity lasers appears to be of crucial
importance for the design, tailoring and optimization of
$Q$-values of lasing microdisk cavities. Such the studies would
require an effective computational method that can deal with both
complex geometry  and variable refraction index in the cavity.

One of the most powerful and versatile numerical techniques often
used in photonic simulation is the finite difference time domain
method (FDTD) \cite{Yee,Li,Sadiku}. A severe disadvantage of this
technique in application to the cavities with small surface
imperfections is that the smooth geometry of the cavity has to be
mapped into a discrete grid with very small lattice constant. This
makes the application of this method to the problem at hand rather
impractical in terms of both computational power and memory.

Another class of computational methods reduces the Helmholtz
equation in the infinite two-dimensional space into contour
integral equations defined at the cavity boundaries. These methods
include the $T$-matrix technique \cite{Waterman,Mishchenko}, the
boundary integral methods \cite{Knipp,Wiersig}, and others
\cite{Boriskina}. These methods are computationally effective and
capable to deal with the cavities of arbitrary geometry. However,
the above methods require the refraction index be constant inside
the cavity boundary.

In the present paper we develop a new, computationally effective,
and numerically stable approach based on the scattering matrix
technique that is capable to deal with \textit{both} arbitrary
complex geometry and inhomogeneous refraction index inside the
cavity. Note that the scattering matrix technique is widely used
in analysis of waveguides\cite{Russian} as well as in quantum
mechanical simulations\cite{Datta}.  This technique was also used
for the analysis of resonant cavities for geometries when the
analytical solution was available\cite{Hentschel}.

The main idea of the method consists of dividing the cavity region
into $N$ narrow concentric rings. At each $i$-th boundary between
the neighboring rings we calculate the scattering matrix
$\mathbf{S^i}$ that relates the states propagating (or decaying)
towards the boundary, with those propagating (or decaying) away of
the boundary. Successively combining the scattering matrixes for
all the boundaries\cite{Russian,Datta},
$\mathbf{S^1}\otimes\ldots\otimes\mathbf{S^{N}}$, we eventually
relate the combined matrix to the total scattering matrix of the
cavity $\mathbf{S}$. In order to calculate the lifetime of the
cavity modes (and, therefore their $Q$-factor) we compute the
Wigner-Smith lifetime matrix\cite{Smith} which, in turn, is
expressed in terms of the total scattering matrix $\mathbf{S}$
\cite{Smith,Mello,Nockel}.

Because at each step we combine only two scattering matrixes, it
is not required to keep track of the solution in the whole space.
This obviously eliminates the need for storing large matrices and
facilitates the computational speed. It is also well known that
the scattering matrix technique (in contrast, for example, to the
transfer matrix technique) is not plagued by numerical
instability, because exponentially growing and decaying evanescent
waves are separated in course of the computation. Note that the
present technique of combining $S$-matrixes is conceptually
similar to the recurrence algorithm for calculating
electromagnetic scattering from a multilayered sphere
\cite{Wu,Johnson}. However, in contrast to these works, the
scattering matrix technique presented here can be applied to the
systems where the refraction index varies as a function of both
radial and angular coordinates.

The paper is organized as follows. In Section \ref{basics} we
develop the scattering matrix technique for disk-shaped cavities.
The results of numerical calculations of resonant states and
$Q$-values of nonideal cavities on the basis of the developed
technique are presented in Section \ref{results}. We consider and
compare two cases, a disk cavity of a constant refraction index
$n$ with side wall imperfection (surface roughness), and, a disk
cavity of an ideal circular shape but with inhomogeneous
refraction index $n=n(r,\varphi)$. The results of numerical
simulation  are analyzed and explained in terms of wave reflection
at a curved dielectric interface combined with the examination of
Poincar\'{e} surfaces of section and Husimi function. Finally, we
present our conclusion in Section \ref{Conclusions}.

\section{The scattering matrix approach}\label{S_martix_section}
\subsection{Formalism}\label{basics}

We consider a two-dimensional cavity with the refraction index $n$
surrounded by air. Because the majority of experiments are
performed only with the lowest transverse mode occupied, we
neglect the transverse ($z$-) dependence of the field and thus
limit ourself to the two-dimensional Helmholtz equation. The
two-dimensional Helmholtz equation for $z$-components of
electromagnetic field is given by
\begin{equation}
\label{Helmholtz} \left(\frac{\partial^2}{\partial
{r^2}}+\frac{1}{r} \frac{\partial}{\partial{r}}+\frac{1}{r^2}
\frac{\partial ^2}{\partial {\varphi ^2}}\right) \Psi (r,\varphi)
+(kn)^2\Psi (r,\varphi)=0,
\end{equation}
where $\Psi=E_z\ (H_z)$ for TM (TE)-modes, and $k$ is the wave
vector in vacuum. Remaining components of the electromagnetic
field can be derived from $E_z\ (H_z)$ in a standard way.
%____________________________________________________________________________________
\begin{figure}[!htp]
\includegraphics[scale=0.5]{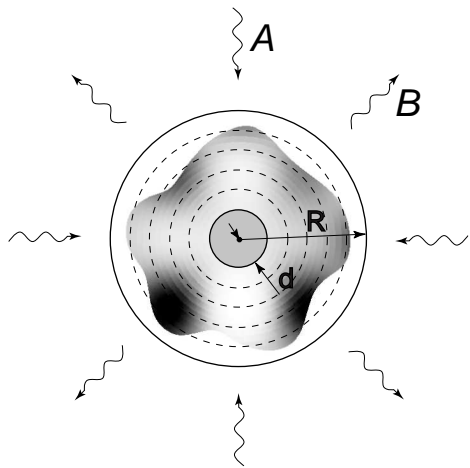}
\caption{Schematic geometry of a cavity with the refraction index
$n$ surrounded by air. The space is divided in three regions. In
the inner ($r<d$) and in the outer regions ($r>R$) the refraction
indexes are constant. In the intermediate region $d<r<R$ the
refraction index $n$ is a function of both $r$ and $\varphi$. The
intermediate region is divided by $N$ narrow concentric rings. In
each ring the refraction coefficient is regarded as a function of
the angle only, $n_i=n_i(\varphi)$. } \label{three_regions}
\end{figure}
%____________________________________________________________________________________

We divide our system in three region, the outer region, $r>R$, the
inner region, $r<d$, and the intermediate region, $d<r<R$, see
Fig. \ref{three_regions}. We choose $R$ and $d$ in such a way that
in the outer and the inner region the refraction indexes are
independent of the coordinate,
 whereas in the intermediate region $n$ is a function of both $r$ and $\varphi$. In the outer region
 the solution to the Helmholtz equation can be written in the form
 \begin{equation}\label{psi_outer}
 \Psi_{out}=\sum_{q=-\infty}^{+\infty}\left(A_q H_q^{(2)}(kr)+B_q H_q^{(1)}(kr)
 \right)e^{iq\varphi},
 \end{equation}
where $H_q^{(1)},H_q^{(2)}$ are the Hankel functions of the first
and second kind of the order $q$ describing respectively incoming
and outgoing waves.

We define the scattering matrix $\mathbf{S}$ in a standard fashion
\cite{Datta,Russian} ,
\begin{equation}\label{scatter_matrix}
B=\mathbf{S}A,
 \end{equation}
where $A, B$ are the column vectors composed of the expansion
coefficients $A_q,B_q$ in Eq. (\ref{psi_outer}). The matrix
element $S_{q'q}=(\mathbf{S})_{q'q}$ gives a probability amplitude
of the scattering from the incoming state $q$ into the outgoing
state $q'$. Because of the requirement of the flux conservation,
the scattering matrix is unitary  \cite{Datta},
\begin{eqnarray}\label{unitarity}
\mathbf{S}\mathbf{S}^\dag=\mathbf{I},
\end{eqnarray} where
$\mathbf{I}$ is the identity matrix. The time reversal invariance
imposes the symmetry requirement upon the scattering matrix
\cite{Datta},
\begin{eqnarray}\label{symmetry}
S_{q'q}=S_{qq'}.
\end{eqnarray}
These two conditions can be used to control numerical results for
the scattering matrix.

%____________________________________________________________________________________
\begin{figure}[!htp]
\includegraphics[scale=0.5]{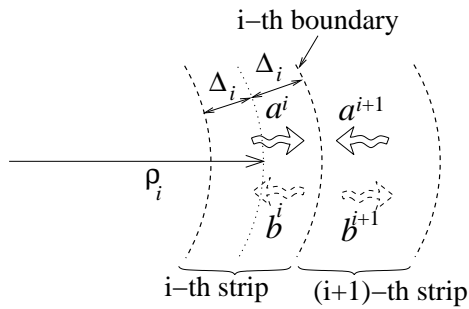}
\caption{The intermediate region is divided by $N$ concentric
rings of the width $2\Delta$; $\rho_i$ is the distance to the
middle of the $i$-th ring. States $a^{i}, a^{i+1}$ propagate (or
decay) towards the $i$-th boundary, whereas states $b^{i},
b^{i+1}$ propagate (or decay) away of this boundary. The $i$-th
boundary is defined as the boundary between the $i$-th and
($i+1$)-th rings. } \label{i-th_boundary}
\end{figure}
%____________________________________________________________________________________
In order to apply the scattering matrix technique we divide the
intermediate region into $N$ narrow concentric rings, see Figs.
\ref{three_regions}, \ref{i-th_boundary}. Within each $i$-th ring
we write down the solution to the Helmholtz equation as a linear
superposition of the states propagating (or decaying) out of the
disk center and the states propagating (or decaying) towards the
disk center (the detailed form of these states will be given in
Section \ref{intermediate_wave_funct}, see Eq.
(\ref{approx_sol_i_ring})). At each $i$-th boundary (defined as a
boundary between the $i$-th and $i+1$-th  rings) we can introduce
the scattering matrix $\mathbf{S^i}$ that relates the states
propagating (or decaying) towards the boundary, $\{a^i_m\}$ and
$\{a^{i+1}_m\}$, with those propagating (or decaying) away of the
boundary, $\{b^i_m\}$ and $\{b^{i+1}_m\}$,
\begin{eqnarray}\label{definition_Si}
\left(\begin{array}{ll}
 b^i\\b^{i+1}
\end{array}\right)=
\mathbf{S^i}
\left(\begin{array}{ll}
 a^i\\a^{i+1}
\end{array}\right), \quad
 1\leq i \leq N-1,
\end{eqnarray}
where $a^i, b^{i}$ are the column vectors composed of the
expansion coefficients $\{a^i_m\}, \{b^i_m\}$, see below,
Eq.(\ref{approx_sol_i_ring}). For the $N$-th boundary between the
last $N$-th ring and the outer region the scattering matrix
$\mathbf{S^N}$ is defined in the form
\begin{eqnarray}\label{definition_SN}
\left(\begin{array}{ll}
 b^N\\B
\end{array}\right)=
\mathbf{S^N} \left(\begin{array}{ll}
 a^N\\A
\end{array}\right).
\end{eqnarray}
In the inner region ($i=0$) the solution to the Helmholtz equation has the form
\begin{equation}\label{psi_in}
 \Psi_{in}=\sum_{q=-\infty}^{+\infty}a^0_q J_q(nkr)e^{iq\varphi},
 \end{equation}
where $J_q$ is the Bessel functions of the order $q$. For the
inner boundary ($i=0$) between the inner region and the first ring
in the intermediate region we define the matrix $\mathbf{S^0}$
according to
\begin{eqnarray}\label{definition_S0}
\left(\begin{array}{ll}
 a^0\\b^1
\end{array}\right)=
\mathbf{S^0} \left(\begin{array}{ll}
 a^0\\a^1
\end{array}\right).
\end{eqnarray}
The brief outline of the derivation and the expressions for the scattering matrixes $\mathbf{S^i}$
are given in Section \ref{Si} and Appendix \ref{Appendix_A}.

The essence of the scattering matrix technique is the successive
combination of the scattering matrixes in the neighboring regions.
For example, combining the scattering matrixes for the $i$-th and
$i+1$-th boundaries, $\mathbf{S^{i}}$ and $\mathbf{S^{i+1}}$, we
obtain the combined scattering matrix $\mathbf{{\tilde{S}}^{i,
i+1}}=\mathbf{S^i}\otimes\mathbf{S^{i+1}}$ that relates the
outgoing and incoming states in the rings $i$ and $i+2$
\cite{Russian,Datta},
\begin{eqnarray}\label{combination}
\left(\begin{array}{ll}
 b^i\\b^{i+2}
\end{array}\right)&=&
\mathbf{{\tilde{S}}^{i, i+1}} \left(\begin{array}{ll}
 a^i\\a^{i+2}
\end{array}\right)\\ \nonumber
\mathbf{{\tilde{S}}^{i, i+1}_{11}}&=&\mathbf{S^{i}_{11}}+\mathbf{S^{i}_{12}}\mathbf{S^{i+1}_{11}}
\left(\mathbf{I}-\mathbf{S^{i}_{22}}\mathbf{S^{i+1}_{11}}\right)^{-1}\mathbf{S^{i}_{21}},
\\ \nonumber
\mathbf{{\tilde{S}}^{i, i+1}_{12}}&=&\mathbf{S^{i}_{12}}\left(\mathbf{I}-\mathbf{S^{i+1}_{11}}
\mathbf{S^{i}_{22}}\right)^{-1}\mathbf{S^{i+1}_{12}}
\\ \nonumber
\mathbf{{\tilde{S}}^{i, i+1}_{21}}&=& \mathbf{S^{i+1}_{21}}
\left(\mathbf{I}-\mathbf{S^{i}_{22}}\mathbf{S^{i+1}_{11}}\right)^{-1}\mathbf{S^{i}_{21}}
\\ \nonumber
\mathbf{{\tilde{S}}^{i, i+1}_{22}}&=&\mathbf{S^{i+1}_{22}}+\mathbf{S^{i+1}_{21}}
\left(\mathbf{I}-\mathbf{S^{i}_{22}}\mathbf{S^{i+1}_{11}}\right)^{-1}
\mathbf{S^{i}_{22}}\mathbf{S^{i+1}_{12}}
\end{eqnarray}
Here and hereafter we use the notation $\mathbf{S_{11}}, \mathbf{S_{12}}, \ldots$ to define the
respective matrix elements of the block matrix $\mathbf{S}$. Combining step by step
%successively
all the scattering matrixes for all the boundaries $0\leq i \leq N$ we numerically obtain the
total combined matrix  $\mathbf{{\tilde{S}}^{0,
N}}=\mathbf{S^0}\otimes\mathbf{S^{1}}\otimes\ldots\mathbf{S^{N}}$ relating the scattering states
in the outer region ($i=N$) and the states in the inner region ($i=0$),
\begin{eqnarray}\label{S_total}
\left(\begin{array}{ll}
 a\\B
\end{array}\right)=
\mathbf{{\tilde{S}}^{0, N}} \left(\begin{array}{ll}
 a\\A
\end{array}\right).
\end{eqnarray}
In order to obtain the scattering matrix $\mathbf{S}$ defined by Eq. (\ref{scatter_matrix}), we
eliminate $a$ from Eq. (\ref{S_total}) and find the relation between $\mathbf{\tilde{S}^{0, N}}$
and $\mathbf{S}$,
\begin{eqnarray}\label{S_total_S}
\mathbf{S}=\mathbf{\tilde{S}^{0, N}_{21}}\left(\mathbf{I}-\mathbf{\tilde{S}^{0,
N}_{11}}\right)^{-1} \mathbf{\tilde{S}^{0, N}_{12}}+\mathbf{\tilde{S}^{0, N}_{22}}.
\end{eqnarray}

To identify the resonant states of an open cavity we introduce the
lifetime matrix (often called as Wigner-Smith time-delay matrix)
\cite{Smith}
\begin{eqnarray} \mathbf{Q}=\frac{i}{c}\,  \frac{d\mathbf{S^\dag}}{dk}\,\mathbf{S}=
-\frac{i}{c}\, \mathbf{S^\dag}\, \frac{d\mathbf{S}}{dk}.
\end{eqnarray}
The diagonal elements of this matrix give a time delay experienced
by the wave incident in $q$-th channel and scattered into all
other channels,
\begin{eqnarray}
\tau_D^q(k)=\mathbf{Q}_{qq}=\frac{i}{c}\, \sum_{q'} \frac{d\mathbf{S^\dag}_{qq'}}{dk}\,
\mathbf{S}_{q'q}.
\end{eqnarray}
The delay time $\tau_D^q(k)$ experienced by a scattering wave is
totally equivalent to the lifetime $\tau=1/2ck^{''}$ of a
quasi-bound state with complex eigenvector $k=k-ik^{''}$
\cite{Nockel}. It is interesting to note that Smith in his
original paper dealing with quantum mechanical scattering
\cite{Smith} chose a letter ``$Q$" to define the lifetime matrix
of a quantum system because of a close analogy to the definition
of a $Q$-value in electromagnetic theory. The total time delay
averaged over all $M$ incoming channels can be expressed in the
form \cite{Mello,Nockel}
\begin{eqnarray}\label{tau_D}
\tau_D(k)=\frac{1}{M}\sum_{q=1}^M\tau_D^q(k)=\frac{1}{M}\,\frac{i}{c}\,
\textrm{Tr}\left(\frac{d\mathbf{S^\dag}}{dk}\,\mathbf{S}\right)=\frac{1}{cM}\sum_{\mu=1}^{M}\frac{d
\theta_\mu }{dk}=\frac{1}{cM}\frac{d \theta}{dk}\, ,
\end{eqnarray}
where $\exp(i\theta_\mu)=\lambda_\mu$ are the eigenvalues of the scattering matrix $\mathbf{S}$,
$\theta=\sum_{\mu=1}^N\theta_\mu$ is the total phase of the determinant of the matrix
$\mathbf{S}$, $\det \mathbf{S}=\prod_{\mu=1}^{M}\lambda_\mu=\exp(i\theta)$.

The resonant states are manifested as peaks in the delay time
whose positions  determine the resonant wavevectors $k_{res}$, and
the heights are related to the $Q$-value of the cavity according
to
\begin{eqnarray}\label{Q_definition}
Q=\omega \tau_D(k_{res}).
\end{eqnarray}

\subsection{Calculation of the wave functions in the intermediate region $d<r<R$}
\label{intermediate_wave_funct}

In the intermediate region the refraction index $n$ depends on
both $r$ and $\varphi$. Therefore, in contrast to the inner and
outer regions, in the intermediate region we can not separate
variables and find an exact analytical solution to the Helmholtz
equation. We can however write down an \textit{approximate}
solution to the Helmholtz equation in each ring. For this purpose
let us look for the solution in the form
$\Psi(r,\varphi)=R(r)\Phi(\varphi)$. Substituting this solution
into Eq. (\ref{Helmholtz}) we obtain
\begin{eqnarray}\label{separated_1}
\frac{r^2}{R(r)}\frac{\partial^2R(r)}{\partial r^2}+
\frac{r}{R(r)}\frac{\partial R(r)}{\partial
r}=-\frac{1}{\Phi(\varphi)}\frac{\partial^2\Phi(\varphi)}{\partial\varphi^2}-k^2n^2(r,\varphi)r^2.
\end{eqnarray}
Let us now assume that each ring with radius $\rho_i$ has a
vanishing width $2\Delta \rightarrow 0$ (see Fig.
\ref{i-th_boundary}). In this case we can regard $r$ as a constant
within each $i$-th ring, $r\approx\rho_i$, with the refraction
index being a function of the angle only
$n(r,\varphi)=n_i(\varphi)$.  In this approximation the variables
in Eq. (\ref{separated_1}) separate such that for $i$-th ring we
can write
\begin{eqnarray}\label{separated_2(a)}
\frac{\partial^2\Phi^i(\varphi)}{\partial \varphi^2} + \left(\zeta^i
+k_i^2(\varphi)\rho_i^2\right)\Phi^i(\varphi)=0
\\
 \label{separated_2(b)}
 \frac{\partial^2R^i(r_i)}{\partial r_i^2}+\frac{\partial R^i(r_i)}{\partial r_i} - \zeta^i R^i(r_i)=0,
\end{eqnarray}
where $\zeta^i$ is a constant (which can be both positive and
negative), and $r_i=r/\rho_i$. The angular function
$\Phi^i(\varphi)$ satisfies the cyclic boundary condition
$\Phi^i(0)=\Phi^i(2\pi)$.  The solution of Eq.
(\ref{separated_2(a)}) thus provides an infinite set of
eigenvalues $\{\zeta^i_m\}$ with the corresponding eigenfunctions
$\Phi^i_m(\varphi)$. Generally, Eq. (\ref{separated_2(a)}) has to
be solved numerically. For a given eigenvalue $\zeta^i_m$ the
solution of Eq. (\ref{separated_2(b)}) for the radial wave
function can be easily written in the analytical form, and the
approximate solution to the Helmholtz equation in the $i$-th ring
(situated to the left to $i$-th boundary) reads
\begin{eqnarray}\label{approx_sol_i_ring}
\Psi_i(r_i,\varphi)=\sum_{m=1}^{\infty} \left( a_m^i
e^{\left(-\frac{1}{2}+i\gamma_m^i\right)\tilde{r}_i} +
 b_m^i
e^{\left(-\frac{1}{2}-i\gamma_m^i\right)\tilde{r}_i} \right) \Phi^i_m(\varphi),
\end{eqnarray}
where $\tilde{r}_i=(r-\rho_i)/\rho_i$ and
$\gamma^i_m=\sqrt{-\frac{1}{4}-\zeta^i_m}$. The states in Eq.
(\ref{approx_sol_i_ring}) are grouped  according to the convention
adopted in the previous subsection \ref{basics}. Namely, the
states propagating to the right towards the $i$-th boundary
($e^{+i\gamma_m^i\tilde{r}_i}$) are described by the coefficients
$\{a_m\}$, whereas the states propagating away from the $i$-th
boundary ($e^{-i\gamma_m^i\tilde{r}_i}$) enter with the
coefficients $\{b_m\}$. Note that if $\gamma^i_m$ becomes
imaginary, $\gamma=i\kappa$, the state propagating towards (away
of) the $i$-th boundary turns into the states decaying towards
(away of) this boundary.

The wave function $\Psi_{i+1}(r_{i+1},\varphi)$ in the ($i+1$)-th
ring (situated to the right to $i$-th boundary) is given by the
similar expression with coefficients $a_m$ and $b_m$ interchanged,
\begin{eqnarray}\label{approx_sol_i+1_ring}
\Psi_{i+1}(r_{i+1},\varphi)=\sum_{m=1}^{\infty} \left( b_m^{i+1}
e^{\left(-\frac{1}{2}+i\gamma_m^{i+1}\right)\tilde{r}_{i+1}} +
 a_m^{i+1}
e^{\left(-\frac{1}{2}-i\gamma_m^{i+1}\right)\tilde{r}_{i+1}} \right) \Phi^{i+1}_m(\varphi),
\end{eqnarray}
  This is because in the ($i+1$)-th ring the states
$e^{+i\gamma_m^{i+1}\tilde{r}_{i+1}}$ propagate (or decay) away of the $i$-th boundary, whereas
the states $e^{-\gamma_m^{i+1}\tilde{r}_{i+1}}$ propagate (or decay) towards the $i$-th boundary.

\subsection{The scattering matrix $\mathbf{S^i}$ at the $i$-th boundary}
\label{Si}

In this section we derive the expression for the scattering matrix
$\mathbf{S^i}$ by matching the wave functions across the $i$-th
boundary. Using the condition of the continuity of the tangential
components of the electric and magnetic fields at the boundary
between two dielectric media, the matching conditions at the
$i$-th boundary (i.e. at the boundary between $i$-th and $i+1$
rings ) read
\begin{eqnarray}\label{matching_condition}
\Psi_i(r,\varphi)&=&\Psi_{i+1}(r,\varphi),\\
\nonumber \frac{1}{\chi_i^2(\varphi)}\frac{\partial \Psi_i(r,\varphi)}{\partial
r}&=&\frac{1}{\chi_{i+1}^2(\varphi)}\frac{\partial \Psi_{i+1}(r,\varphi)}{\partial r},
\end{eqnarray}
where $\chi_i^2(\varphi)=1$ for TM modes, and
$\chi_i^2(\varphi)=k^2n_i^2(\varphi)$ for TE modes.

In order to derive the expression for the scattering matrix
$\mathbf{S^i}$ in the intermediate region ($1\leq i \leq N-1$) we
substitute the wave functions Eqs. (\ref{approx_sol_i_ring}),
(\ref{approx_sol_i+1_ring}) into the boundary conditions Eq.
(\ref{matching_condition}). Multiplying the obtained equations by
$\left(\Phi^{i}_m(\varphi)\right)^*$ and integrating over the
angle using the conditions of the orthogonality
$\int_0^{2\pi}d\varphi\,
\left(\Phi^{i}_m(\varphi)\right)^*\Phi^{i}_{m'}(\varphi)=\delta_{mm'}$
we arrive to two infinite systems of equations for the
coefficients $a_m^{i},a_m^{i+1},b_m^{i},b_m^{i+1}$. After some
straightforward algebra these systems of equations are reduced to
the form prescribed by Eq. (\ref{definition_Si}) with the
following result
\begin{eqnarray}\label{Si_general_expression}
\mathbf{S^{i}}=\mathbf{\Lambda}\mathbf{K}\mathbf{A}^{-1}\mathbf{B}
 \mathbf{K}\mathbf{\Lambda}^{-1}.
\end{eqnarray}
The scattering matrixes $\mathbf{S^{0}}, \mathbf{S^{N}}$ (for
inner $i=0$ and outer $i=N$ boundaries  respectively) are derived
in a similar fashion. The expression for $\mathbf{S^{i}}$ given by
Eq. (\ref{Si_general_expression}) holds for all the boundaries
$0\leq i\leq N$. A particular form of the matrixes
$\mathbf{\Lambda},\mathbf{K},\mathbf{A},\mathbf{B}$ is different
for three distinct cases, namely, \emph{(a)} 0-th boundary (the
boundary between the inner region ($i=0$) and the first ring $i=1$
in the intermediate region); \textit{(b)} $i$-th boundary,
$1<i<N-1$, (the boundary between $i$-th and $i+1$-th rings in the
intermediate region), and \textit{(c)} $N$-th boundary (the
boundary between the last ring $i=N$ in the intermediate region
and the outer region ($i=N+1$)). The corresponding expressions for
these three cases are given in Appendix, Eqs.
(\ref{i=0})-(\ref{i=N}).

\section{Nonideal microdisk cavities}\label{results}

In this Section we apply the scattering matrix method  to the
calculation of resonant states and $Q$-values of nonideal
microdisk cavities with (a) side wall imperfections and (b)
circular cavities with inhomogeneous refraction index
$n=n(r,\varphi)$. In order to validate present method, we have
also performed numerical calculations for structures where the
analytical solution was available. This includes, for example, an
annular billiard consisting of a dielectric disk placed inside a
larger disk with some displacement of the disk center
\cite{Hentschel}, as well as an ideal circular disk displaced from
the origin of coordinate system. In the latter case, the positions
of the resonant states and $Q$-values are obviously independent of
the choice of the coordinate system. However, from computational
point of view this case is not simpler than that of a cavity of an
arbitrary shape, because the displacement from the origin lifts
the radial symmetry and makes the separation of variables
impossible. As an additional tool to validate the numerical
solution we use Eqs. (\ref{unitarity}),(\ref{symmetry}) to control
the unitarity and symmetry of the scattering matrix.

\subsection{Ideal circular cavity}

Let us first briefly analyze the resonant states and $Q$-values of
an ideal circular cavity with the radius $R$ and the refraction
index $n$. In this case the scattering matrix can be easily
written in analytical form. Employing the matching conditions Eq.
(\ref{matching_condition}) between the wave function in the outer
region $r>R$, Eq. (\ref{psi_outer}), and the wave function inside
the disk given by the Bessel functions $J_q(nkr)$ for $r<R$, Eq.
(\ref{psi_in}), we arrive to the expression for the scattering
matrix in the form \cite{Hentschel}
\begin{eqnarray}\label{S_ideal_disk}
S_{qq'}=
\frac{
H_q^{(2)'}(kr)-\xi
\left[
J_q'(nkr)/J_q(nkr)
\right]H_q^{(2)}(kr)
}
{
H_q^{(1)'}(kr)-\xi\left[
J_q'(nkr)/J_q(nkr)
\right]H_q^{(1)}(kr)
}\delta_{qq'},
\end{eqnarray}
with $\xi=n$ $(1/n)$ for TM (TE) modes.  Derivatives are taken
over the full arguments in the brackets. Resonant states of an
ideal cavity can be inferred from the scattering matrix Eq.
(\ref{S_ideal_disk}) using Eq. (\ref{tau_D}).

Each resonant states of an open disk is characterized by two wave
numbers, $q$ and $m$. These two numbers are directly related to
the corresponding numbers of the closed resonator of the same
radius $R$. The index $m$ is a radial wave number and it is
related to the number of nodes of the field components in the
radial direction inside the disk. The index $q$ is called an
angular (or azimuthal) wave number because of the analogy to
quantum mechanics where the angular momentum is given by
$L_{QM}=\hbar q$. Equating the quantum and classical angular
momenta ($L_{Clas}=pR\sin\chi, p=\hbar nk$) we find the relation
between the angular wave number and the angle of incidence $\chi$
in a classical ray picture \cite{Hentschel}
\begin{equation}
q=nkR\, \sin\chi .\label{semiclassics}
\end{equation}

Here we are mostly interested in the whispering gallery modes with
high $Q$-values for which the angle of incidence  is larger than
the angle of total internal reflection, $\chi>\chi_c$
($\sin\chi_c=1/n$). For such angles of incidence, the transmission
probability $T$ of an electromagnetic wave incident  on a curved
interface of radius $\rho$ is small, $T\ll 1$. For the case when
the radius of curvature is much larger than the wavelength,
$kn\rho \gg 1$, (which applies to majority of cavities), the
transmission probability reads \cite{Snyder}
\begin{eqnarray}\label{Eq_T}
T = |T_{F}| \exp{\left[-\frac{2}{3}\frac{nk\rho}{\sin^2(\chi)}
\left(\cos^2\chi_{c}-\cos^2\chi\right)^{3/2}\right]},
\end{eqnarray}
where $T_{F}$ is the classical Fresnel transmission coefficient
for an electromagnetic wave incident on a flat surface.
%____________________________________________________________________________________
\begin{figure}[!htp]
\includegraphics[scale=0.4]{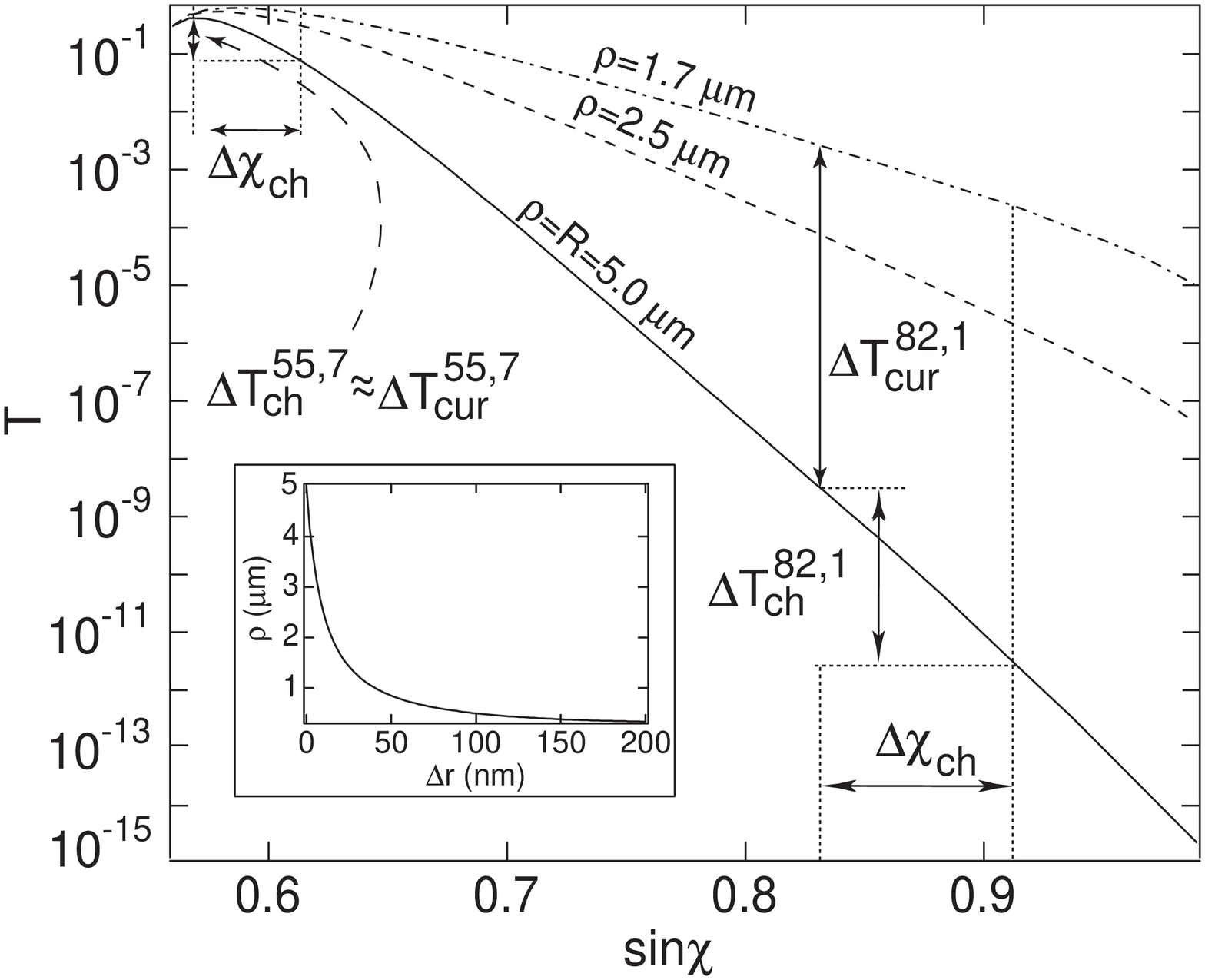}
\caption{Transmission coefficient $T$ of a locally plane wave
incident on a curved surface with the radii of curvature $\rho$
 as a function of the incidence angle
$\chi$ calculated from Eq. (\ref{Eq_T}). The angle of total
internal reflection $\sin\chi_c=0.56$ (corresponding to $n=1.8$).
The inset shows the dependence of  the average radius of local
curvature due to boundary imperfections, $\rho$, subject to
$\Delta r$ for the present model of surface roughness.} \label{T}
\end{figure}
%____________________________________________________________________________________

Figure \ref{T} illustrates that $T$ decreases exponentially as the
difference $\chi-\chi_c$ grows. The $Q$-value of the whispering
gallery mode $q$ in a cavity of the radius $R$ is related to the
transmission probability $T$, Eq. (\ref{Eq_T}), by the relation
\cite{Hentschel_rapid}
\begin{eqnarray} Q=\frac{2nkR
\cos\chi}{T},
\end{eqnarray}
 where the classical incidence angle $\chi$  is related to mode number $q$ by
Eq. (\ref{semiclassics}), and $T\ll 1$.

\subsection{Nonideal cavities with (a) surface roughness and
(b) inhomogeneous refraction index}

In this Section we present the results of numerical calculations
of resonant states and $Q$-values of nonideal cavities. We
consider separately two cases, (a), a disk cavity of a constant
refraction  index $n$ but with side wall imperfection (surface
roughness), and, (b), a disk cavity of an ideal circular shape but
with inhomogeneous refraction index $n=n(r,\varphi)$.

\begin{figure}[!htp]
\includegraphics[scale=0.6]{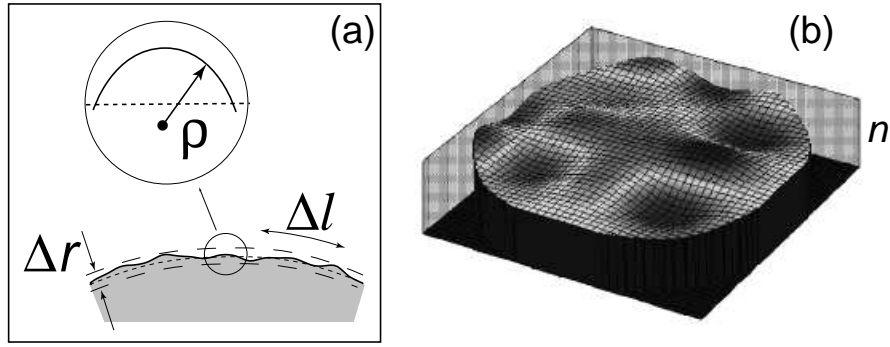}
\caption{Examples of nonideal cavities studied in the present
paper with (a) surface roughness and (b) inhomogeneous refraction
index. (a) Radius of the disk $R=5\mu$m, $n=1.8$, surface
roughness $\Delta r=100$ nm. (b) $R=5\mu$m, $\langle n\rangle
=1.8$, $\Delta n=5\%$.} \label{examples}
\end{figure}

Various studies indicate that a typical size of the side-wall
imperfections can vary in the range of 5-300 nm (representing a
variation of the order of $\sim$0.05-1\% of the cavity radius).
\cite{Fujita,Gayral,Seassal,Polson2002}. An exact experimental
shape of the cavity-air interface is however not available. We
thus model the interface shape as a superposition of random
Gaussian deviations from an ideal circle of radius $R$ with a
maximal amplitude $\Delta r/2$ and a characteristic distance
between the deviation maxima $\Delta l\sim 2\pi R/50 $. In a
similar fashion we model the inhomogeneity of the diffraction
index in the cavity, where a parameter $\Delta n$ characterizes a
mean deviation of the refraction index $n$ from its average value
$\langle n\rangle=1.8$. The variation of the refraction index $n$
can be caused by different factors including the presence of
quantum well/wires/dots forming an active medium of the cavity,
the local field intensity dependence $n=n(I)$, and other factors.
Examples of typical structures under investigation are shown in
Fig. \ref{examples}.

%____________________________________________________________________________________
%____________________________________________________________________________________

\begin{figure}[!htp]
\includegraphics[scale=0.35]{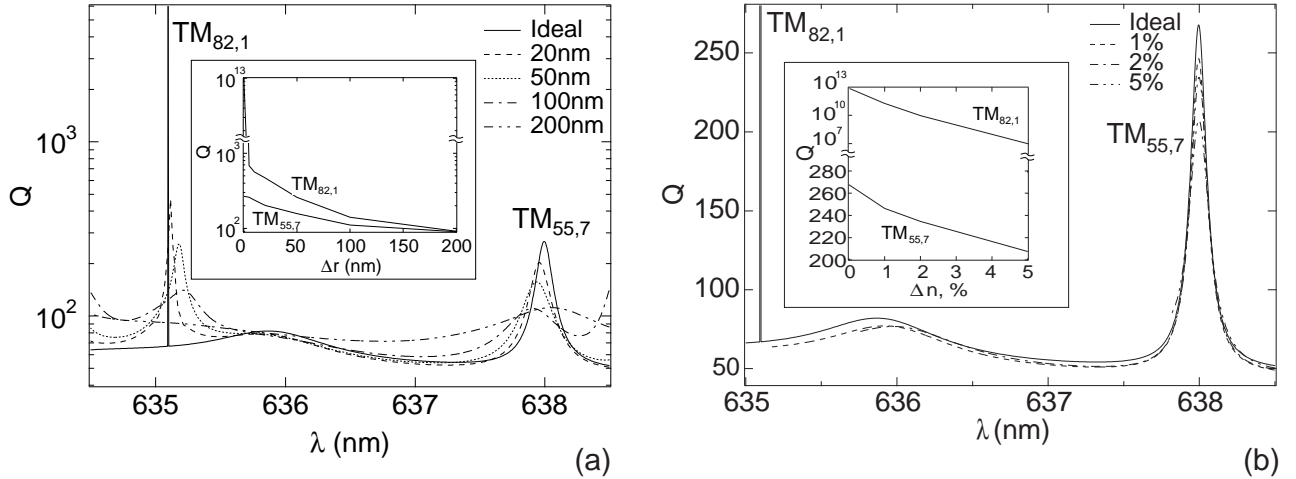}
\caption{Dependencies $Q=Q(\lambda)$  for two representative modes
TM$_{82,1}$ and TM$_{55,7}$ for the cases of (a) different surface
roughness $\Delta r$  and (b) different refraction index
inhomogeneities. The values of $\Delta r$ and $\Delta n$ are
indicated in Figs. (a) and (b) respectively; $R=5 \mu$m, $n=1.8$
(a), $\langle n\rangle =1.8$ (b). Note that in the case (b) the
resonances shift when $\Delta n$ varies. For the sake of clearness
we plot all the resonances centered around their maxima of the
corresponding ideal disk (i. e. $\Delta n=0$). The broadening of
the high-$Q$ resonance TM$_{55,7}$ is not discernible on the scale
of the figure for all the values of $\Delta n$. Insets in Figs.
(a) and (b) show the dependencies $Q=Q(\Delta r)$ and $Q=Q(\Delta
n)$ respectively.} \label{spectra}
\end{figure}

Figure \ref{spectra} shows calculated $Q$-values of the disk
resonant cavity for  different surface roughnesses $\Delta r$ and
the refraction index inhomogeneity $\Delta n$  in some
representative wavelength interval. Note that we have studied a
number of different resonances and all of them showed the same
trends described below. Besides, here and hereafter we concentrate
only on TM modes of the cavity, because TE modes  exhibit similar
features. The calculated dependencies of the $Q$-values on
 $\Delta r$ and $\Delta n$ are summarized in the insets to Fig. \ref{spectra}.

%____________________________________________________________________________________
%____________________________________________________________________________________

Let us first concentrate on the low-$Q$ state TM$_{55,7}$ ($q=55,
m=7$). A decrease of the both surface roughness $\Delta r$ and the
refractive index inhomogeneity $\Delta n$ causes graduate and
rather slow decrease of the $Q$-value of this state as shown in
the insets to Fig. \ref{spectra}. This behavior is typical for all
other low-$Q$ states. In contrast, the high-$Q$ resonances exhibit
very different and rather striking behavior. Namely, these
resonances show  a dramatic decrease of their $Q$-values even for
very small values of the surface roughness $\Delta
r\lesssim\lambda/50$. At the same time, the $Q$-values of the
cavity decrease much more slowly when the refractive index
inhomogeneity $\Delta n$ increases. For example, let us choose
$\Delta r=20$nm and $\Delta n=5\%$. For these values of $\Delta r$
and $\Delta n$ the $Q$-value of the low-$Q$ state TM$_{55,7}$
drops by the same factor of $\sim 1.3$, decreasing from $Q\approx
270 $ to $Q\approx 205$. In contrast, for the very same surface
roughness $\Delta r$, the $Q$-value of a high-$Q$ state
TM$_{82,1}$ drops by the factor of $\sim  10^{11}$ decreasing from
its value $Q\approx 4\cdot 10^{13}$ for an ideal disk to $Q\approx
260$. At the same time, for the above value of $\Delta n=5\%$, the
$Q$-value of this resonance decreases to the value of $Q\approx
1.3\cdot 10^7$, which corresponds to the drop by the factor $\sim
10^4$. (Note that for the case of an ideal cavity the high-$Q$
resonances are so narrow that the numerical resolution does not
allow a reliable estimation of their exact values. In this case we
therefore use Eq. (\ref{T}) to estimate their $Q$-values.)

\subsection{Discussion}

\begin{figure}[!htp]
\includegraphics[scale=0.6]{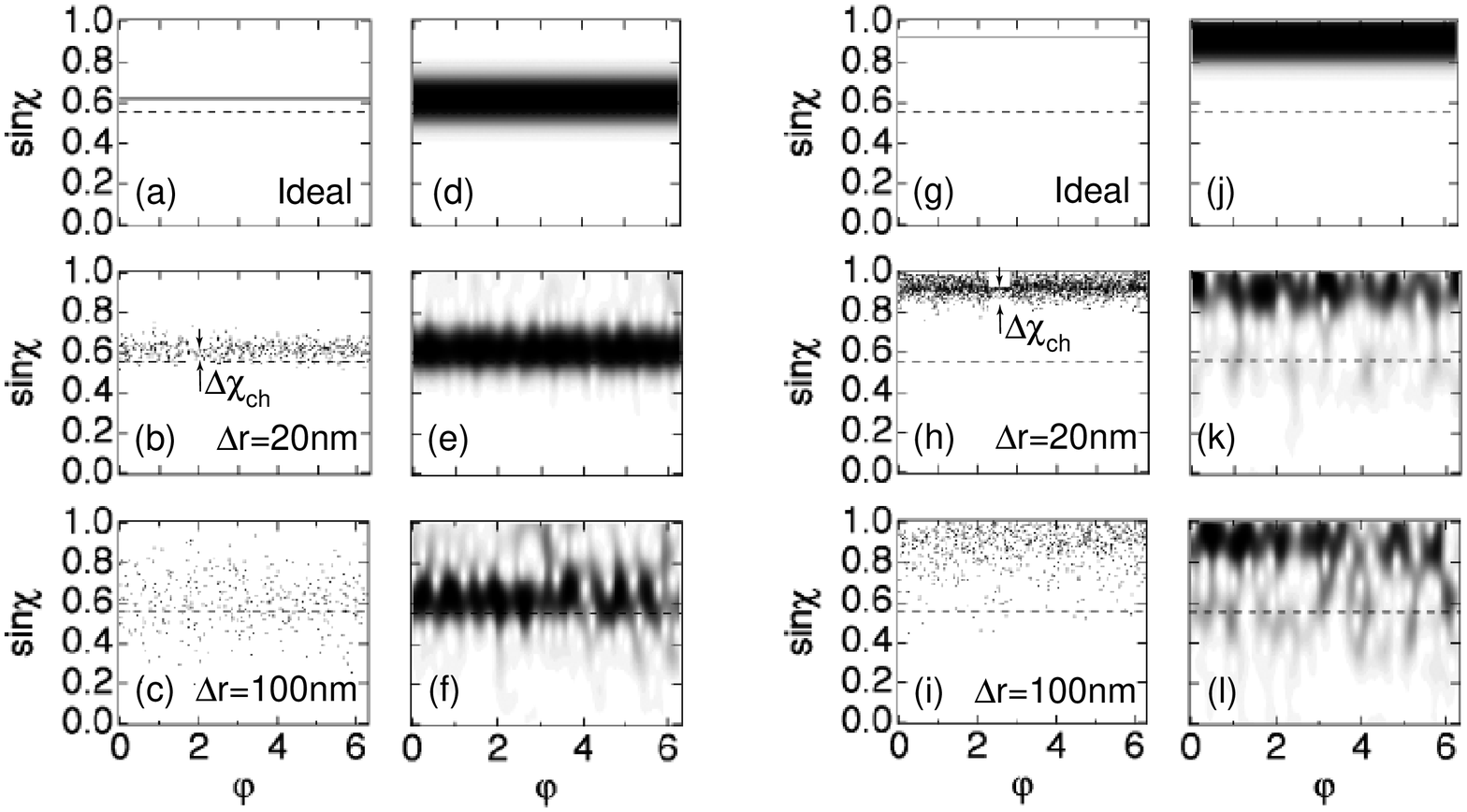}
\caption{ Poincar\'{e} surfaces of section for geometrical rays
corresponding to the states $q=55$ (a)-(c) and $q=82$ (g)-(i) for
the cavity with the surface roughness $\Delta r = 0,
20\textrm{nm}, 100\textrm{nm}$.
 The Husimi distributions for the states TM$_{55,7}$
(d)-(f) and TM$_{82,1}$ (j)-(l) for the same values of $\Delta r$
as in the corresponding Poincar\'{e} SoS. $\Delta
\chi_{\textrm{ch}}$ indicate the broadening of the phase space due
to transition to the chaotic dynamics. Dashed lines show the angle
of total internal reflection $\chi_c$.} \label{PSoS_Husimi}
\end{figure}

In the previous Section we found that the surface roughness
$\Delta r$ and refraction index inhomogeneity $\Delta n$ that
produce similar degradation of low-$Q$ states, cause strikingly
different effect on high-$Q$ resonances. In order to understand
these features, we shall combine Poincar\'{e} surface of section
and Husimi function methods  with an analysis of ray reflection at
a curved dielectric interface. The Poincar\'{e} surface of section
(SoS) is a powerful tool visualizing the phase space for a
classical ray dynamics in cavities \cite{Nockel,Stone}. We
concentrate on the surface section of the phase space along the
cavity boundary, $r\in \textrm{surf}$. For a given resonant state
with an angular number $q$, the corresponding ray is launched with
the angle $\chi_0=\arcsin\left(q/nkR\right)$ according to Eq.
(\ref{semiclassics}). Each reflection at the boundary
(characterizing by the polar angle $\varphi$, and the angle of
incidence $\chi$), corresponds to a single point in the plot. The
number of bounces for a given angle of incidence $\chi_0$ is
chosen in such a way that the total path of the ray does not
exceed the one extracted from the numerically calculated $Q$-value
for the corresponding resonance, $L=c\tau_D=Q/(kn)$. Figures
\ref{PSoS_Husimi} (a)-(c), (g)-(i) show a Poincar\'{e} surfaces of
section (SoS) for the geometrical rays corresponding to the states
TM$_{55,7}$, TM$_{82,1}$ for different values of the surface
roughness $\Delta r$. For an ideal circular disk ($\Delta r=0$)
the Poincar\'{e} SoS are obviously straight lines corresponding to
a constant angle of incidence $\chi=\chi_0$. Figures
\ref{PSoS_Husimi} (b)-(c), (h)-(i) demonstrate that initially
regular dynamics of an ideal cavity transforms into a chaotic one
even for a cavity with maximum roughness $\Delta r \lesssim 20$nm.
In Figs. \ref{PSoS_Husimi} (b)-(c), (h)-(i) $\Delta
\chi_{\textrm{ch}}$ approximately indicates the broadening of the
phase space due to transition to the chaotic dynamics. An
important observation is that for a given surface roughness
$\Delta r$ the broadening of the phase space is independent of
angular mode $q$, i.e. it is the same for low- and high-$Q$
states.

%____________________________________________________________________________________
%____________________________________________________________________________________

\begin{figure}[!htp]
\includegraphics[scale=0.7]{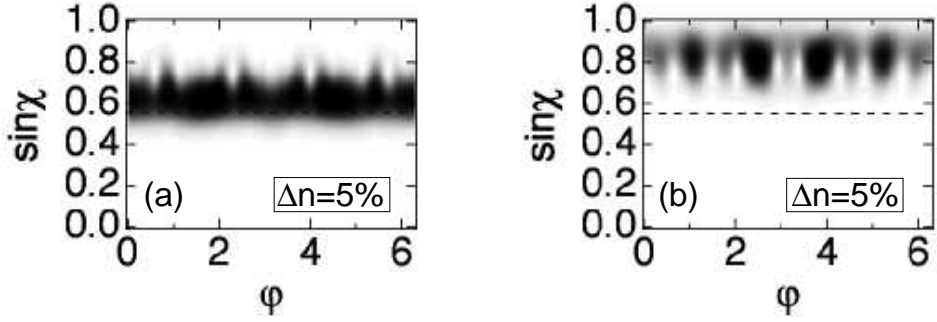}
\caption{Illustrative examples of intensity distribution $E_z$ for
the resonant state TM$_{55,7}$ in cavities with $\Delta r=0$ (a),
$\Delta r=20 \textrm{nm}$ (b), $\Delta r=20 \textrm{nm}$ (c); $R=5
\mu$m, $n=1.8$. Dashed lines indicate boundaries of the cavity.}
\label{E_distributions}
\end{figure}

We complement classical Poincar\'{e} SoS by the Husimi function
analysis \cite{Husimi,Nockel,Stone}. The Husimi function (often
called also Husimi distributions) $H(\varphi,\chi )$ represents a
quantum (wave) analog to a classical Poincar\'{e} SoS. It is
defined as a projection of a given cavity mode $\Psi{}(r\in
\textrm{surf},\varphi)$ taken at the surface of cavity
 into a Gaussian wave packet
$\Phi(\varphi';\varphi,\chi)$ impinging the cavity boundary with
the coordinate $\varphi$ at the angle $\chi$,
\begin{equation}\label{husimi}
H(\varphi ,\chi ) = \int_0^{2\pi } {d\varphi ' \Psi(r\in
\textrm{surf},\varphi' )\Phi(\varphi';\varphi,\chi)},
\end{equation}
where the minimum-uncertainty wave packet centered around
$\varphi,\chi$ with the dispersion in position $\sqrt{\sigma/2}$
is given by
\begin{equation}\label{gwp}
\Phi(\varphi';\varphi,\chi) = (\pi \sigma )^{ - \frac{1}{4}}
\sum_l {\exp\left[{ - \frac{1}{{2\sigma }}(\varphi'- \varphi  +
2\pi l )^2 - ik\sin \chi (\varphi  + 2\pi l)}\right] },
\end{equation}
where we have chosen $\sigma=\sqrt{2}/k$. The Husimi
distributions, Figs. \ref{PSoS_Husimi} (d)-(f), (j)-(l), exhibit
the same trends as the classical Poincar\'{e} SoS. Indeed,
broadening the phase space with the increasing the surface
roughness $\Delta r$ for the Husimi functions is the same as the
corresponding broadening $\Delta \chi_{\textrm{ch}}$ in the
Poincar\'{e} SoS. (Illustrative examples of the wave functions in
cavities for different surface roughness $\Delta n$ are shown in
Fig. \ref{E_distributions}).

%____________________________________________________________________________________
%____________________________________________________________________________________

%____________________________________________________________________________________
\begin{figure}[!htp]
\includegraphics[scale=0.7]{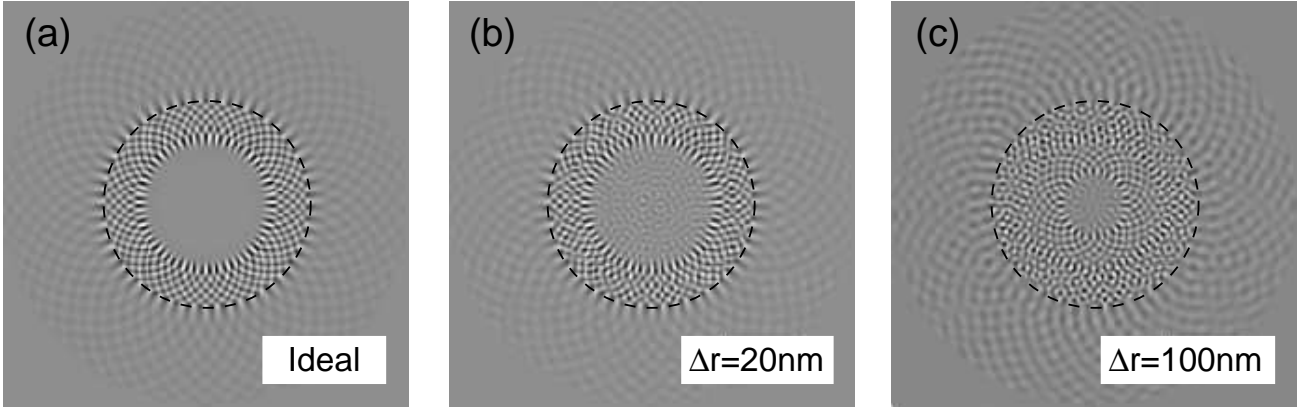}
\caption{The Husimi distributions for the states TM$_{55,7}$ (a)
 and TM$_{82,1}$ (b) for the cavity with the refraction index
inhomogeneity $\Delta n =5\%$.} \label{PSoS_Husimi_n}
\end{figure}
%____________________________________________________________________________________

Figure \ref{PSoS_Husimi_n} shows the Husimi distributions for a
circular cavity with an inhomogeneous refraction index.  The
variation of the refraction index $\Delta n=5\%$ is chosen in such
a way that the degradation of the $Q$-value for the low-$Q$
resonance TM$_{55,7}$ is the same as the one for the case of
surface roughness $\Delta r = 20\textrm{nm}$ shown in Fig.
\ref{PSoS_Husimi}. As expected, the broadening of the Husimi
distribution due to increase of $\Delta n$ is of the same order as
for the corresponding values $\Delta r$ (compare Figs.
\ref{PSoS_Husimi} and \ref{PSoS_Husimi_n}).

According to Eq (\ref{T}),  one can expect an increase of the
transmission coefficient (and therefore decrease of the $Q$-value
of the cavity) due to the broadening of the phase space $\Delta
\chi_{\textrm{ch}}$, because the incidence angle $\chi$
effectively moves closer to the angle of the total internal
reflection $\chi_c$. $\Delta T_{\textrm{ch}}$ in Fig. \ref{T}
indicates the estimated increase of the transmission coefficient
due to the broadening of the phase space, $\Delta
\chi_{\textrm{ch}}$, as extracted from the Poincar\'{e} SoS  for
the $\Delta r=20$nm and $\Delta n=5\%$. For the low-$Q$ resonance
TM$_{55,7}$ this corresponds to the decrease of the $Q$-value by
the factor of $\Delta Q_{\textrm{ch}}\sim\Delta
T^{-1}_{\textrm{ch}}\approx 1.5$, which is consistent with the
calculated decrease of the low-$Q$ resonances.

For the case of high-$Q$ resonance TM$_{82.1}$ the estimated
decrease of the $Q$-factor is $\Delta Q_{\textrm{ch}}\sim\Delta
T^{-1}_{\textrm{ch}}\approx 10^3\sim 10^4$ (see Fig. \ref{T}),
which is consistent with the calculated decrease of this resonance
for the case of the inhomogeneous refraction index only. (Note
that because of a rather approximate definition of $\Delta
\chi_{\textrm{ch}}$ we can give only very rough estimation of the
factor $\Delta T_{\textrm{ch}}$.) On the contrary, for the case of
high-$Q$ resonances in the presence of surface imperfections, this
estimated value of $\Delta Q_{\textrm{ch}}$ is in many orders of
magnitude smaller that the actual calculated decrease of the
$Q$-factor (given by factor of $\approx10^{11}$, see Fig.
\ref{spectra}).

To explain the rapid degradation of high-$Q$ resonances, we
concentrate on another aspect of the wave dynamics. Namely, the
imperfections at the surface boundary effectively introduce a
local radius of a surface curvature $\rho$  that is distinct from
the disk radius $R$ (see illustration in Fig. \ref{examples}). One
may thus expect that with the presence of the local surface
curvature, the total transmission coefficient will be determined
by the averaged value of $\rho$ rather than by the disk radius
$R$. The dependence of $\rho$ on surface roughness $\Delta r$ for
the present model of surface imperfections is shown in the inset
to Fig. \ref{T}. Figure \ref{T} demonstrates that  the reduction
of the local radius of curvature from $5\mu$m (ideal disk) to
$1.7\mu$m ($\Delta r = 20$nm) causes an increase of the
transmission coefficient by $\Delta T_{\textrm{cur}}\approx 10^8$.
This estimate, combined with the estimate based on the change of
$\Delta T_{\textrm{ch}}$ is fully consistent with the actual
computed decrease of the $Q$-factor shown in Fig. \ref{spectra}.
We thus conclude that the main mechanism responsible for the rapid
degradation of high-$Q$ resonances in non-ideal cavities is the
enhanced radiative decay through the curved surface because the
effective local radius (given by the surface roughness) is smaller
that the disk radius $R$.

In contrast, the degradation of low-$Q$ resonances (as well as
high-$Q$ resonances in the case of inhomogeneous refraction index
only), is mostly related to the broadening of the phase space
caused by the transition to the chaotic dynamics. It should be
noted however that both factors (broadening of the phase space and
the enhancement of the transmission due to decrease of the
effective radius of curvature) may play a comparable role in
degradation of the low-$Q$ whispering-gallery resonances in the
presence of surface roughness.

It is interesting to note that an analogues degradation of
high-$Q$ modes was recently found in hexagonal-shaped
microcavities, where the modes were strongly influenced by
roundings of the corners even when the characteristic length scale
(the local radius of curvature) was one order of magnitude smaller
than the wavelength \cite{Wiersig_PRA}. It is worth mentioning
that one often assumes that long-lived high-$Q$ resonances in
idealized cavities (e.g. in ideal disks, hexagons, etc.) are not
important for potential application in optical communication or
laser devices\cite{Hentschel,Wiersig} because of their extremely
narrow width. Our simulations demonstrate that it is not the case,
because in real structures
 the $Q$-values of these
resonances becomes comparable to those of intermediate-$Q$
resonances already for small or moderate surface roughness of
$\Delta r\sim 10-50$ nm.

\section{Conclusions}\label{Conclusions}
In the present paper we develop a new, computationally effective,
and numerically stable approach based on the scattering matrix
($S$-matrix) technique that is capable to deal with \textit{both}
arbitrary complex geometry and inhomogeneous refraction index
inside the two-dimensional cavity. The derivation is based on the
separation of the cavity region into $N$ narrow concentric rings
and calculation of the $S$-matrix at every boundary between the
rings. The total $S$-matrix is obtained in a recursive way by
successive combination of the scattering matrixes for all the
boundaries. In order to calculate the lifetime of the cavity modes
(and, therefore their $Q$-factors) we compute the Wigner-Smith
time delay-matrix which, in turn, is expressed in terms of the
total scattering matrix.

We apply the developed algorithm  to the calculation of resonant
states and $Q$-values of nonideal microdisk cavities with (a) side
wall imperfections and (b) circular cavities with inhomogeneous
refraction index $n=n(r,\varphi)$. We find that the surface
roughness $\Delta r$ and refraction index inhomogeneity $\Delta n$
that produce similar degradation of low-$Q$ states, cause
strikingly different effect on high-$Q$ resonances. In
particularly, in the case of inhomogeneous refraction index the
increase of $\Delta n$ causes rather graduate decrease of the
$Q$-value of high-$Q$ resonances. In contrast, in the presence of
surface roughness even small imperfections ($\Delta r
\lesssim\lambda/50$) can lead to a drastic degradation of high-$Q$
cavity modes by many orders of magnitude.

In order to understand these features, we combine Poincar\'{e}
surface of section and Husimi function methods  with an analysis
of ray reflection at a curved dielectric interface. We argue that
the main mechanism responsible for the rapid degradation of
high-$Q$ resonances in non-ideal cavities with the surface
roughness is the enhanced radiative decay through the curved
surface because the effective local radius (given by the surface
roughness) is smaller that the disk radius $R$. In contrast, the
degradation of low-$Q$ resonances (as well as high-$Q$ resonances
in the case of inhomogeneous refraction index only), is mostly
related to the broadening of the phase space caused by the
transition to the chaotic dynamics.

\begin{acknowledgments}
We thank Olle Ingan\"as for stimulating discussions that initiated
this work. We are also grateful to Sayan Mukherjee and especially
to Stanley Miklavcic for many useful discussions and
conversations. We appreciate correspondence with Jan Wiersig.
A.I.R. acknowledges financial support from SI and KVA.
\end{acknowledgments}

\appendix

\section{Expressions for the matrixes $\mathbf{S^i}$}\label{Appendix_A}

In this appendix we present the expressions for the matrixes
$\mathbf{\Lambda},\mathbf{K},\mathbf{A},\mathbf{B}$ entering Eq.
(\ref{Si_general_expression}) for the scattering matrix
$\mathbf{S^i}$ relating incoming and outgoing states at the $i$-th
boundary. We distinguish three different cases as specified below.

 \emph{(a) 0-th
boundary (the boundary between the inner region ($i=0$) and the
first ring $i=1$ in the intermediate region)},

\begin{eqnarray}\label{i=0}
%_____________________________________________________________________
&&\mathbf{\Lambda_{11}}=\mathbf{I},\quad
(\mathbf{\Lambda_{22}})_{mj}=e^{-\frac{1}{2}\Delta_{1}}\delta_{mj},
\quad \mathbf{\Lambda_{12}}=
\mathbf{\Lambda_{21}}=0,\\
%_____________________________________________________________________
\nonumber
&&\mathbf{K_{11}}=\mathbf{I},\quad
(\mathbf{K_{22}})_{mj}=e^{i\gamma_m\Delta_1}\delta_{mj},
\quad \mathbf{K_{12}}=
\mathbf{K_{21}}=0,\\
%_____________________________________________________________________
\nonumber
&&\mathbf{A}=
\left(\begin{array}{lcc}
 \mathbf{0}&\mathbf{V^{0,1}}\\
 -\mathbf{J'}&\mathbf{U^{0,1}P^{1}}
\end{array}\right), \quad
\mathbf{B}=\left(\begin{array}{cc}
 \mathbf{J}&-\mathbf{V^{0,1}}\\
 \mathbf{0}&-\mathbf{U^{0,1}Q^{1}}
\end{array}\right)\\
%_____________________________________________________________________
\nonumber &&(\mathbf{J})_{mj}=J_m(n_0kd)\, \delta_{mj},\quad
(\mathbf{J'})_{mj}=J_m'(n_0kd)\, \delta_{mj},
\\
%_____________________________________________________________________
\nonumber
&&(\mathbf{V^{0,1}})_{mj}=\int_0^{2\pi}e^{-im\varphi}\Phi^{1}_j(\varphi)\,d\varphi,
\quad (\mathbf{U^{0,1}})_{mj}=\frac{1}{n_0k\rho_1}\int_0^{2\pi}
\frac{\chi_0^2(\varphi)}{\chi_{1}^2(\varphi)}
e^{-im\varphi}\Phi^{1}_j(\varphi)\,d\varphi
%_____________________________________________________________________
\end{eqnarray}
\textit{(b) $i$-th boundary, $1<i<N-1$, (the boundary between
$i$-th and $i+1$-th rings in the intermediate region)}
\begin{eqnarray}
\label{1<i<N-1}
%_____________________________________________________________________
&&(\mathbf{\Lambda_{11}})_{mj}=e^{\frac{1}{2}\Delta_i}\delta_{mj},\quad
(\mathbf{\Lambda_{22}})_{mj}=e^{-\frac{1}{2}\Delta_{i+1}}\delta_{mj},
\quad \mathbf{\Lambda_{12}}=
\mathbf{\Lambda_{21}}=0,\quad
(\mathbf{K})_{mj}=e^{i\gamma_m\Delta_i}\delta_{mj}
\quad \\
%_____________________________________________________________________
\nonumber
&&\mathbf{A}=
\left(\begin{array}{lcc}
 -\mathbf{I}&\mathbf{V^{i,i+1}}\\
 -\mathbf{Q^i}&\mathbf{U^{i,i+1}P^{i+1}}
\end{array}\right), \quad
\mathbf{B}=\left(\begin{array}{cc}
 \mathbf{I}&-\mathbf{V^{i,i+1}}\\
 \mathbf{P^i}&-\mathbf{U^{i,i+1}Q^{i+1}}
\end{array}\right)\\
%%_____________________________________________________________________
%\nonumber
%&&(\mathbf{P^i})_{mj}=\left(-\frac{1}{2}+i\gamma_m^i\right)\delta_{mj},\quad
%(\mathbf{Q^i})_{mj}=\left(-\frac{1}{2}-i\gamma_m^i\right)\delta_{mj}
%\\
%_____________________________________________________________________
\nonumber
&&(\mathbf{V^{i,i+1}})_{mj}=\int_0^{2\pi}
\left(\Phi^i_m(\varphi)\right)^*
\Phi^{i+1}_j(\varphi)\,d\varphi, \quad
(\mathbf{U^{i,i+1}})_{mj}=\frac{\rho_i}{\rho_{i+1}}\int_0^{2\pi}
\frac{\chi_i^2(\varphi)}{\chi_{i+1}^2(\varphi)}
\left(\Phi^i_m(\varphi)\right)^*\Phi^{i+1}_j(\varphi)\,d\varphi
%_____________________________________________________________________
\end{eqnarray}

\textit{(c) $N$-th boundary (the boundary between the last ring
$i=N$ in the intermediate region and the outer region ($i=N+1$))}

\begin{eqnarray}\label{i=N}
%_____________________________________________________________________
&&(\mathbf{\Lambda_{11}})_{mj}=e^{\frac{1}{2}\Delta_{N}}\delta_{mj},\quad
\mathbf{\Lambda_{22}}=\mathbf{I},
\quad \mathbf{\Lambda_{12}}=
\mathbf{\Lambda_{21}}=0\\
%_____________________________________________________________________
\nonumber
&&(\mathbf{K_{11}})_{mj}=e^{i\gamma_m\Delta_N}\delta_{mj},\quad
\mathbf{K_{22}}=\mathbf{I},
\quad \mathbf{K_{12}}=
\mathbf{K_{21}}=0,\\
%_____________________________________________________________________
\nonumber
&&\mathbf{A}=
\left(\begin{array}{lcc}
 -\mathbf{I}&\mathbf{V^{N,N+1}H^{(1)}}\\
 -\mathbf{Q^N}&\mathbf{U^{N,N+1}{H^{(1)}}'}
\end{array}\right), \quad
\mathbf{B}=\left(\begin{array}{cc}
 \mathbf{I}&-\mathbf{V^{N,N+1}H^{(2)}}\\
 \mathbf{P^N}&-\mathbf{U^{N,N+1}{H^{(2)}}'}
\end{array}\right)\\
%_____________________________________________________________________
\nonumber &&(\mathbf{H^{(1,2)}})_{mj}=H^{(1,2)}(kR)\,
\delta_{mj},\quad
(\mathbf{{{H^{(1,2)}_{m}}'}})_{mj}={{H^{(1,2)}_{m}}'}(kR)\,
\delta_{mj},
\\
%_____________________________________________________________________
\nonumber &&(\mathbf{V^{N,N+1}})_{mj}=\int_0^{2\pi}
\left(\Phi^{N}_m(\varphi)\right)^* e^{ij\varphi}\,d\varphi, \quad
(\mathbf{U^{N,N+1}})_{mj}=k\rho_N\int_0^{2\pi}
\frac{\chi_N^2(\varphi)}{\chi_{N+1}^2(\varphi)}
\left(\Phi^{1}_j(\varphi)\right)^* e^{ij\varphi} \,d\varphi
%_____________________________________________________________________
\end{eqnarray}
In Eqs. (\ref{i=0})-(\ref{i=N}) the matrixes $\mathbf{Q^i}, \mathbf{P^i}$ are defined according to
\begin{eqnarray}
%_____________________________________________________________________
\nonumber &&(\mathbf{P^i})_{mj}=\left(-\frac{1}{2}+i\gamma_m^i\right)\delta_{mj},\quad
(\mathbf{Q^i})_{mj}=\left(-\frac{1}{2}-i\gamma_m^i\right)\delta_{mj}, \quad 1\leq i\leq N.
\end{eqnarray}

 ${J^{}_{m}}, {H^{(1,2)}_{m}}$, and ${J^{}_{m}}', {H^{(1,2)}_{m}}'$ are the
Bessel and Hankel functions and their derivatives,  and $\Delta_i=\Delta/\rho_i$.


\begin{thebibliography}{99}

\bibitem{Yamamoto}Y. Yamamoto and R. E. Slusher,
``Optical processes in microcavities", Physics Today, June 1993,
p.66.

\bibitem{Nockel}
J. U. N\"ockel and R. K. Chang, ``2-d microcavities: Theory and
Experiments", in \textit{Cavity-Enhanced Spectroscopies},  R.D.
van Zee and J.P.Looney, eds., (Vol. 40 of ``Experimental Methods
in the Physical Sciences", Academic Press, San Diego, 2002), pp.
185-226.

\bibitem{Hill_Benner} S. C. Hill and R. E. Benner,
``Morphology-dependent Resonances", in \textit{Optical Effects
Associated with Small Particles}, P. W. Barber and R. K. Chang,
eds. (Vol. 1 of ``Advanced Series in Applied Physics", World
Scientific, Singapore, 1989).

%----------- dielectric lasers------------------------


\bibitem{Slusher_1992} S. L. McCall, A. F. J. Levi, R. E. Slusher,
 S. J. Pearton, and R. A. Logan, ``Whispering-gallery mode microdisk lasers"
Appl. Phys. Lett. \textbf{60}, 289 (1992).

\bibitem{Slusher_1993} R. E. Slusher,  A. F. J. Levi, U. Mohideen, S. L. McCall,
S. J. Pearton, and R. A. Logan, ``Threshold characteristics of
semiconductor microdisk lasers", Appl. Phys. Lett. \textbf{60},
289 (1992).

\bibitem{Fujita} M. Fujita, K. Inoshita, and T. Bata,
``Room temperature continuous wave lasing characteristics of
GaInAsP/InP microdisk injection laser", Electronic Lett.,
\textbf{34}, 278-279 (1998).

\bibitem{Gayral} B. Gayral, J. M. G\'{e}rard, A. Lema\^{i}tre,
C. Dupuis, L. Manin, and J. L. Pelouard, ``High-$Q$ wet etched
GaAS microdisks containing InAs quantum boxes", Appl. Phys. Lett.
\textbf{75}, 1908-1910 (1999).

\bibitem{Seassal} C. Seassal, X. Letartre, J. Brault, M. Gendry, P. Pottier,
P. Viktorovitch, O. Piquet, P. Blondy, D. Cros, O. Marty,  ``InAs
wuantum wires in InP-based microdiscs: Mode identification and
continuos wave room temperature laser operation", J. Appl. Phys.,
\textbf{88}, 6170-6174 (2000).

%----------- polymeric lasers------------------------

\bibitem{Berggren} A. Dodabalapur, M. Berggren, R. E. Slusher, Z. Bao, A.
Timko, P. Schiortino, E. Laskowski, H. E Katz, and O. Nalamasu,
``Resonators and materials for organic lasers based on energy
transfer" IEEE Journal of selected topics in quantum electronics,
\textbf{4}, 67-74 (1998).

\bibitem{Inganas}M. Theander, T. Granlund, D. M. Johanson,
A. Ruseckas, V. Sundstr\"om, M. R. Andersson, and O. Ingan\"as,
``Lasing in a microcavity with an oriented liquid-crystalline
polyfluorene copolymer as active layer", Adv. Mater. \textbf{13},
323-37 (2001).

\bibitem{Polson2002} R. C. Polson, Z. Vardeny, and D. A. Chinn,
``Multiple resonances in microdisk lasers of $\pi$-conjugated
polymers", Appl. Phys. Lett. \textbf{81}, 1561-1563 (2002).

%----------- Numerical methods, S-matrix------------------------

\bibitem{Yee} K. S. Yee, ``Numerical solution of initial
boundary-value problems involving Maxwell's equations in isotropic
media", IEEE Trans. Ant. Prop., \textbf{AP-14}, 302-307 (1996).

\bibitem{Li}B.-J. Li and P.-L. Liu, ``Analysis of far-field
patterns of microdisk resonators by the finite-difference
time-domain method", IEEE J. Quantum Electron. \textbf{33}, 1489
(1997).

\bibitem{Sadiku} M. N. O. Sadiku, \textit{Numerical Techniques in
Electromagnetics}, (CRC Press, Boca Raton, 2001).

\bibitem{Waterman}P. C. Waterman, Symmetry, unitarity and
geometry in electromagnetic scattering, Phys. Rev. D \textbf{3},
825-839 (1971).

\bibitem{Mishchenko} M. I. Mishchenko, L. D. Travis, and A. A.
Lacis, \textit{Scattering, Absorption, and Emission of Light by
Small Particles}, (Campridge University Press, Cambridge, 2002).

\bibitem{Knipp} P. A. Knipp and T. L. Reinecke, ``Boundary-element method
for the calculation of the electronic states in semiconductor
nanostructures", Phys. Rev. B \textbf{54}, 1880-1891 (1996).

\bibitem{Wiersig}J. Wiersig, J. Opt. A: Pure Appl. Opt.
``Boundary element method for resonances in dielectric
micrcavities", \textbf{5}, 53-60 (2003).

\bibitem{Boriskina} S. V. Boriskina, T. M. Benson, P. Sewell, and
A. I. Nosich, ``Highly efficient design of spectrally engineered
whispering-gallery-mode microlaser resonators", Opt. and Quant.
Electr. \textbf{35}, 545-559 (2003).

\bibitem{Russian} V. V. Nikolsky, T. I. Nikolskaya, \textit{Decomposition approach to the
problems of electrodynamics} (Nauka, Moskow, 1983), (in Russian).

\bibitem{Datta}S. Datta, \textit{ Electronic Transport in Mesoscopic Systems}
(Cambridge University Press, Cambridge, 1995).

\bibitem{Hentschel}M. Hentschel and K. Richter, ``Quantum chaos in optical system:
The annular billiar", Phys. Rev. E \textbf{66}, 056207 1-13
(2002).

\bibitem{Smith}F. T. Smith, ``Lifetime matrix in collision theory", Phys. Rev, \textbf{118}, 349 (1960).

\bibitem{Mello}M. Bauer, P. A. Mello, and K. W. McVoy, ``Time delay in nuclear reactions",
 Z. Physik A \textbf{293}, 151 (1979).


\bibitem{Wu} Z. S. Wu and Y. P. Y. Wang, ``Electromagnetic
scattering for multilayered sphere: recursive algorithms", Radio
Sci. \textbf{26}, 1393-1401 (1991).

\bibitem{Johnson} B. R. Johnson, ``Light scattering by a multilayer
sphere", Applied Optics \textbf{35}, 3286-3296 (1996).
%___________________________________________________________





\bibitem{Snyder}A. V. Snyder and J. D. Love,  ``Reflection at a curved dielectric interface --
Electromagnetic tunneling",IEEE\ Trans.\ Microwave.\ Theor.\
Techn. \textbf{MTT-23}, 134-141 (1975).

\bibitem{Hentschel_rapid}M. Hentschel and H. Schomerus, ``Fresnel laws at
curved dielectric interfaces of microresonators", Phys.\ Rev.\ E.
\textbf{65}, 045603 1-4 (R) (2002).

\bibitem{Stone}A. D. Stone, ``Wave-chaotic optical resonators and lasers",
 Physica Scripta \textbf{T90}, 448 (2001) (Proceedings of
the Nobel Symposium "Quantum Chaos 2000").

\bibitem{Husimi} B. Crespi, G. Perez, and S. J. Chang, ``Quantum
Poincar\'{e} sections for two-dimentional billiards", Phys. Rev. E
\textbf{47}, 986-991 (1993).

\bibitem{Wiersig_PRA} J. Wiersig, private communication; J. Wiersig,
``Hexagonal dielectric resonators and microcrystal lasers", Phys.
Rev. A \textbf{67} 023807 1-12 (2003).


\end{thebibliography}
\end{document}